\documentclass[preprint,3p]{elsarticle}

\usepackage{hyperref,subcaption,amsmath,booktabs,color}

\journal{Physics Letters B}

\begin{document}

  \begin{frontmatter}

    \title{Erratum to "Measurement of the $e^+e^-\to\pi^+\pi^-$ cross section \\ between 600 and 900 MeV using initial state radiation"}
    
    \author{
      \begin{small}
\begin{center}
M.~Ablikim$^{1}$, M.~N.~Achasov$^{10,c}$, P.~Adlarson$^{67}$, S. ~Ahmed$^{15}$, M.~Albrecht$^{4}$, R.~Aliberti$^{28}$, A.~Amoroso$^{66A,66C}$, Q.~An$^{63,50}$, X.~H.~Bai$^{57}$, Y.~Bai$^{49}$
% ...
\\
\vspace{0.2cm}
(BESIII Collaboration)\\
\vspace{0.2cm} {\it
$^{1}$ Institute of High Energy Physics, Beijing 100049, People's Republic of China\\
% ...
$^{4}$ Bochum Ruhr-University, D-44780 Bochum, Germany\\
% ...
$^{10}$ G.I. Budker Institute of Nuclear Physics SB RAS (BINP), Novosibirsk 630090, Russia\\
% ...
$^{15}$ Helmholtz Institute Mainz, Johann-Joachim-Becher-Weg 45, D-55099 Mainz, Germany\\
% ...
$^{28}$ Johannes Gutenberg University of Mainz, Johann-Joachim-Becher-Weg 45, D-55099 Mainz, Germany\\
% ...
$^{49}$ Southeast University, Nanjing 211100, People's Republic of China\\
$^{50}$ State Key Laboratory of Particle Detection and Electronics, Beijing 100049, Hefei 230026, People's Republic of China\\
% ...
$^{57}$ University of Jinan, Jinan 250022, People's Republic of China\\
% ...
$^{63}$ University of Science and Technology of China, Hefei 230026, People's Republic of China\\
% ...
$^{66}$ University of Turin and INFN, INFN, I-10125, Turin, Italy\\
$^{67}$ Uppsala University, Box 516, SE-75120 Uppsala, Sweden\\
% ...
$^{c}$ Also at the Novosibirsk State University, Novosibirsk, 630090, Russia\\
% ...
}\end{center}

\vspace{0.4cm}
\end{small}

    }
    
    \begin{abstract}
      In Ref.~\cite{Ablikim:2015orh} the BESIII collaboration published a cross section measurement of the process $e^+e^-\to \pi^+ \pi^-$ in the energy range between 600 and 900\,MeV. In this erratum, we report a corrected evaluation of the statistical errors in terms of a fully propagated covariance matrix. The correction also yields a reduced statistical uncertainty for the hadronic vacuum polarization contribution to the anomalous magnetic moment of the muon, which now reads as $a_\mu^{\pi\pi\mathrm{, LO}}(600 - 900\,\mathrm{MeV}) = (368.2 \pm 1.5_{\rm stat} \pm 3.3_{\rm syst})\times 10^{-10}$. The central values of the cross section measurement and of $a_\mu^{\pi\pi\mathrm{, LO}}$, as well as the systematic uncertainties remain unchanged.
    \end{abstract}


  \end{frontmatter}


\section{Introduction}
Previously, we reported~\cite{Ablikim:2015orh} a measurement of the cross section $\sigma^\mathrm{bare}(e^+e^-\to\pi^+\pi^-)$ and the pion form factor $|F_\pi|^2$ in the energy range between 600\,MeV and 900\,MeV. As pointed out in Refs.~\cite{Colangelo:2018mtw} and \cite{Davier:2019can}, there exists a difference between the statistical uncertainties of the tabulated cross section of Ref.~\cite{Ablikim:2015orh} and the covariance matrix, which is documented as a supplemental material to the publication. Furthermore, when including the covariance matrix, it is not possible to reproduce the fit of the form factor presented in Ref.~\cite{Ablikim:2015orh}.
    
In scrutinizing the published analysis, we realized that the covariance matrix had been provided at the level of the event yield, not, as claimed, at the level of the cross section. The same matrix is erroneously used in the calculation of $a_\mu^{\pi\pi\mathrm{, LO}}$. At the same time, the statistical errors in Tab.~4 of Ref.~\cite{Ablikim:2015orh} are taken from the event yield prior to unfolding and are propagated to the level of the cross section and form factor, respectively, producing two different statistical uncertainties for the results.

In this work, the statistical uncertainties are reevaluated and an updated value of the uncertainty of the two-pion contribution to the hadronic vacuum polarization contribution of the anomalous magnetic moment of the muon, $a_\mu^{\pi\pi\mathrm{, LO}}$, is calculated.

\section{Reevaluation of the Statistical Covariance Matrix}
The covariance matrix results from the unfolding procedure, which is applied at the level of the event yield to compensate for mass resolution effects of the detector. The underlying algorithm of the procedure is based on singular value decomposition~\cite{Hocker:1995kb}. The covariance matrix needs to be propagated according to generalized Gaussian error propagation to correctly reflect the statistical correlations of the cross section and the form factor, respectively.

In an initial state radiation (ISR) measurement, the dressed cross section $\sigma^\text{dressed}(e^+e^-\to \pi^+\pi^-)$ is calculated from the unfolded event yield $N_{\rm unf}$ of $\pi^+\pi^-\gamma_{\rm ISR}$ events according to
 \begin{equation}\label{eq:dxs}
  \sigma^\text{dressed}(e^+e^-\to \pi^+\pi^-) = \frac {N_{\rm unf}} {\varepsilon^{\pi\pi}_{\rm global} \cdot \mathcal{L}_\text{int} \cdot H(s,s') \cdot (1+\delta^{\pi\pi\gamma}_{\text{FSR}}) 
  } \quad ,
 \end{equation}
where $\varepsilon^{\pi\pi}_{\rm global}$ is the reconstruction efficiency, $\mathcal{L}_\text{int}$ is the integrated luminosity, and $H(s,s')$ is the radiator function, where the implementation of Ref.~\cite{Rodrigo:2001kf} is considered. The correction $(1+\delta^{\pi\pi\gamma}_{\text{FSR}})$ denotes the final state radiation (FSR) corrections on the level of radiative $\pi^+\pi^-\gamma$ events\footnote{In Eq.~1 of Ref.~\cite{Ablikim:2015orh}, the factor $(1+\delta^{\pi\pi}_{\rm FSR})$ should be read as $\left[\frac{1+\eta(s)\frac{\alpha}{\pi}}{1+\delta^{\pi\pi\gamma}_{\rm FSR}}\right]$, contrary to the description in Section~6.3 therein.}. 

The {\it bare} cross section is obtained from the dressed cross section by applying mass-dependent corrections for vacuum polarization $\delta_\text{VP}$~\cite{Jegerlehner:1985gq} and by adding back effects of FSR on the level of the non-radiative $\pi^+\pi^-$ cross sections as parametrized within scalar QED in the Schwinger term $1+\eta(s')\frac{\alpha}{\pi}$~\cite{Schwinger:1989ix}. The final formula for the bare cross section reads as:
 \begin{equation}\label{eq:bxs}
  \sigma^\text{bare}(e^+e^-\to \pi^+\pi^-(\gamma_\text{FSR}))= \sigma^\text{dressed} (e^+e^-\to \pi^+\pi^-)\frac{1+ \eta(s')\frac{\alpha}{\pi}}{\delta_\text{VP}(s')}
  \quad ,
 \end{equation}
where $s'$ denotes the two-pion invariant mass squared. 

Since all the above mentioned values remain unchanged compared to the original work~\cite{Ablikim:2015orh}, the central value of the cross section does not change.

The covariance matrix obtained from the unfolding procedure is propagated taking into account Eqs.~\ref{eq:dxs} and~\ref{eq:bxs} to the level of the bare cross section. It is, assuming no correlations between the contributing quantities, thus, given by
  \begin{equation}\label{eq:errprop}
    C^{\sigma^\text{bare}} = \sum_{k\in \{ N,\varepsilon,\mathcal{L}_\text{int},H,(1+\delta^{\pi\pi\gamma}_{\text{FSR}}) \}} \left(J^{\text{T}}\right)^k C^kJ^k \quad ,
  \end{equation}
with $J^k_{ij}=\frac{\partial\sigma^\text{bare}_i}{\partial k_j}$ being the Jacobian matrix of the bare cross section with respect to the contribution $k$ to the statistical uncertainty, according to generalized Gaussian error propagation.

Since the time integrated luminosity is a single scalar value, its covariance matrix is simply given by the squared statistical uncertainty of the time integrated luminosity: $C^{\mathcal{L}_\text{int}}=\left(\Delta\mathcal{L}_\text{int}\right)^2$.

It is assumed that the reconstruction efficiency, the time integrated luminosity, the radiator function, as well as the final state radiation correction term are completely uncorrelated. The respective diagonal elements of the covariance matrices are given by the square of the uncertainties. The contribution of the Schwinger correction term is neglected, since as a QED calculation, it is assumed to be exact. In the original work, the uncertainty of the vacuum polarization effect is considered to be purely systematic. Hence, it is also neglected in the calculation of the statistical covariance matrix.

\begin{figure}[htb]
  \centering
  \includegraphics[trim=0 1.5em 0 1em, clip, width=.45\textwidth]{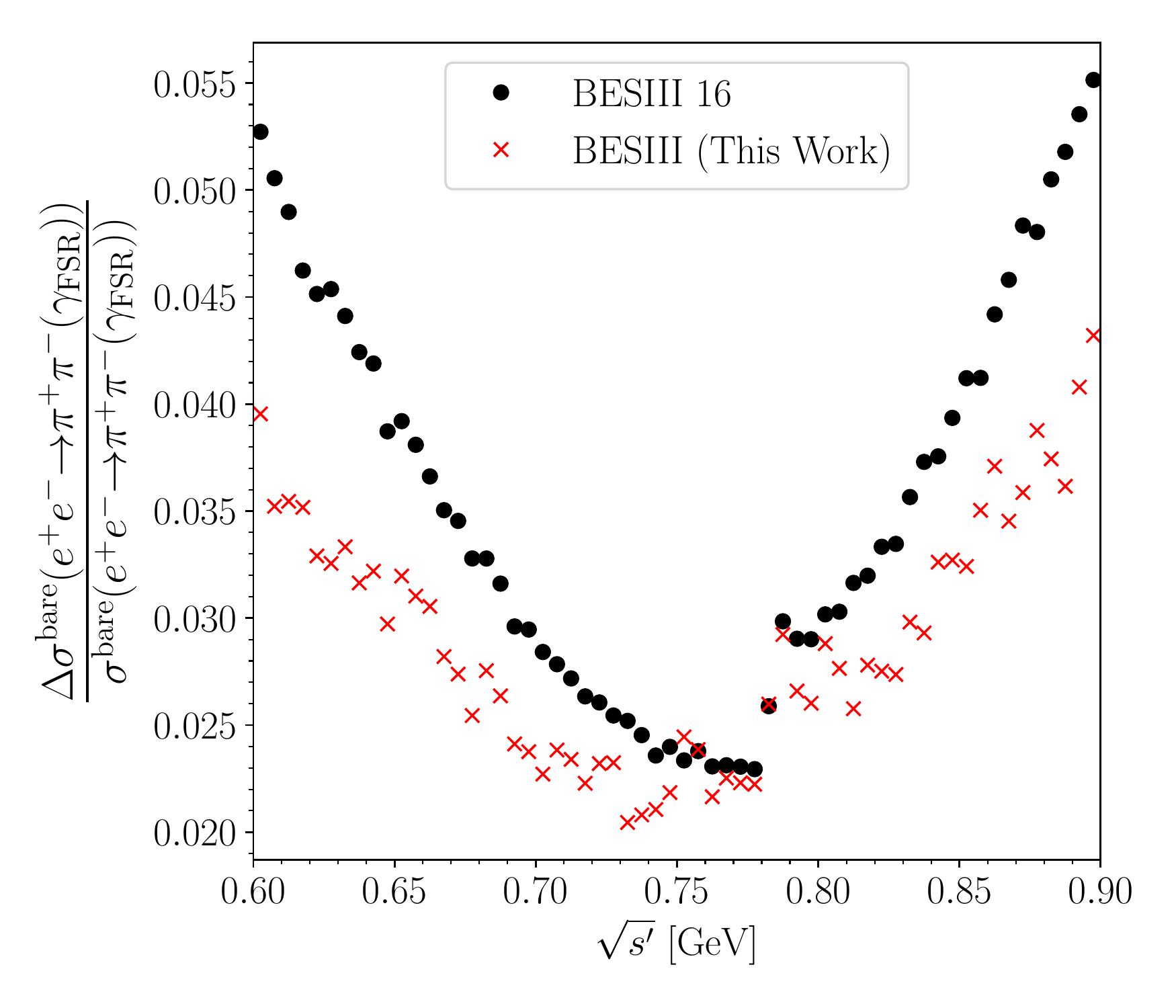}
  \caption{\label{fig:xs_ff}(Color online) Relative uncertainty of the bare cross section $\sigma^\text{bare}(e^+e^+\to\pi^+\pi^-(\gamma_\text{FSR}))$ of this work (red crosses) compared to the results of Ref.~\cite{Ablikim:2015orh} (black circles). The uncertainties of the cross section of this work are the square roots of the diagonal elements of the matrix.}
\end{figure}

In the original publication, the error propagation of the covariance matrix had not been carried out properly. As a result, the statistical uncertainties of the published cross section do not reflect the information of the unfolding. Figure~\ref{fig:xs_ff} shows a comparison of the relative statistical errors of the bare cross sections calculated as the diagonal uncertainties of this work (red crosses) and the uncertainties published in Ref.~\cite{Ablikim:2015orh} (black circles). The values of the diagonal errors are listed in Tab.~\ref{tab:xsec_FF}.

It must be stressed that only the statistical uncertainties of the measurements of $\sigma^\text{bare}(e^+e^+\to\pi^+\pi^-(\gamma_\text{FSR}))$ and of $|F_\pi|^2$ have been reevaluated. Thus, the systematic uncertainty of 0.9\,\% evaluated in Ref.~\cite{Ablikim:2015orh} is unchanged.

The BESIII collaboration has approved new data taking at 3.773\,GeV in 2021-2022, aiming at a total data set of 20\,fb$^{-1}$~\cite{Ablikim:2019hff}. In addition to a significant reduction of the statistical uncertainty, the new data will also allow for the alternative normalization scheme for $\sigma^\text{bare}(e^+e^+\to\pi^+\pi^-(\gamma_\text{FSR}))$, discussed in Eq.~3 of Ref.~\cite{Ablikim:2015orh}, in which the dominating systematic uncertainties cancel. A total uncertainty of 0.6\% can be expected.

\begin{table*}[htb]
  \centering
  \caption{\label{tab:xsec_FF}Results for the bare cross section $\sigma^\text{bare}_{\pi^+\pi^-}$ and the pion form factor together with their statistical uncertainties. The systematical uncertainties are given by 0.9\,\%\cite{Ablikim:2015orh}.}
  \small
  \begin{tabular}{rcc|rcc}
    \toprule 
    $\sqrt{s'}$ [MeV] & $\sigma^\text{bare}_{\pi^+\pi^-(\gamma_\text{FSR})}$ [nb] & $|F_\pi|^2$ & $\sqrt{s'}$ [MeV] & $\sigma^\text{bare}_{\pi^+\pi^-(\gamma_\text{FSR})}$ & $|F_\pi|^2$ \\ \midrule
    602.5 & 288.3 $\pm$ 11.4 & 6.9 $\pm$ 0.3 & 752.5 & 1276.1 $\pm$ 31.2 & 41.8 $\pm$ 1.0 \\
    607.5 & 306.6 $\pm$ 10.8 & 7.4 $\pm$ 0.3 & 757.5 & 1315.9 $\pm$ 31.4 & 43.6 $\pm$ 1.0 \\
    612.5 & 332.8 $\pm$ 11.8 & 8.2 $\pm$ 0.3 & 762.5 & 1339.3 $\pm$ 29.0 & 44.8 $\pm$ 1.0 \\
    617.5 & 352.5 $\pm$ 12.4 & 8.7 $\pm$ 0.3 & 767.5 & 1331.9 $\pm$ 30.0 & 45.0 $\pm$ 1.0 \\
    622.5 & 367.7 $\pm$ 12.1 & 9.2 $\pm$ 0.3 & 772.5 & 1327.0 $\pm$ 29.6 & 45.2 $\pm$ 1.0 \\
    627.5 & 390.1 $\pm$ 12.7 & 9.8 $\pm$ 0.3 & 777.5 & 1272.7 $\pm$ 28.3 & 43.7 $\pm$ 1.0 \\
    632.5 & 408.0 $\pm$ 13.6 & 10.4 $\pm$ 0.3 & 782.5 & 1031.5 $\pm$ 26.8 & 37.1 $\pm$ 1.0 \\
    637.5 & 426.6 $\pm$ 13.5 & 11.0 $\pm$ 0.3 & 787.5 & 810.7 $\pm$ 23.7 & 30.3 $\pm$ 0.9 \\
    642.5 & 453.5 $\pm$ 14.6 & 11.8 $\pm$ 0.4 & 792.5 & 819.7 $\pm$ 21.8 & 30.6 $\pm$ 0.8 \\
    647.5 & 477.7 $\pm$ 14.2 & 12.5 $\pm$ 0.4 & 797.5 & 803.1 $\pm$ 20.9 & 30.1 $\pm$ 0.8 \\
    652.5 & 497.4 $\pm$ 15.9 & 13.2 $\pm$ 0.4 & 802.5 & 732.4 $\pm$ 21.1 & 27.7 $\pm$ 0.8 \\
    657.5 & 509.2 $\pm$ 15.8 & 13.6 $\pm$ 0.4 & 807.5 & 679.9 $\pm$ 18.8 & 25.9 $\pm$ 0.7 \\
    662.5 & 543.4 $\pm$ 16.6 & 14.7 $\pm$ 0.4 & 812.5 & 663.6 $\pm$ 17.1 & 25.5 $\pm$ 0.7 \\
    667.5 & 585.0 $\pm$ 16.5 & 16.0 $\pm$ 0.4 & 817.5 & 622.2 $\pm$ 17.3 & 24.1 $\pm$ 0.7 \\
    672.5 & 642.7 $\pm$ 17.6 & 17.7 $\pm$ 0.5 & 822.5 & 585.0 $\pm$ 16.1 & 22.9 $\pm$ 0.6 \\
    677.5 & 640.5 $\pm$ 16.3 & 17.8 $\pm$ 0.5 & 827.5 & 540.8 $\pm$ 14.8 & 21.4 $\pm$ 0.6 \\
    682.5 & 668.0 $\pm$ 18.4 & 18.8 $\pm$ 0.5 & 832.5 & 496.4 $\pm$ 14.8 & 19.8 $\pm$ 0.6 \\
    687.5 & 724.4 $\pm$ 19.1 & 20.6 $\pm$ 0.5 & 837.5 & 450.4 $\pm$ 13.2 & 18.1 $\pm$ 0.5 \\
    692.5 & 783.5 $\pm$ 18.9 & 22.5 $\pm$ 0.5 & 842.5 & 404.7 $\pm$ 13.2 & 16.4 $\pm$ 0.5 \\
    697.5 & 858.6 $\pm$ 20.4 & 24.9 $\pm$ 0.6 & 847.5 & 391.3 $\pm$ 12.8 & 16.0 $\pm$ 0.5 \\
    702.5 & 893.8 $\pm$ 20.3 & 26.2 $\pm$ 0.6 & 852.5 & 364.0 $\pm$ 11.8 & 15.0 $\pm$ 0.5 \\
    707.5 & 897.8 $\pm$ 21.4 & 26.6 $\pm$ 0.6 & 857.5 & 339.6 $\pm$ 11.9 & 14.2 $\pm$ 0.5 \\
    712.5 & 978.6 $\pm$ 22.9 & 29.3 $\pm$ 0.7 & 862.5 & 310.0 $\pm$ 11.5 & 13.0 $\pm$ 0.5 \\
    717.5 & 1059.1 $\pm$ 23.6 & 32.0 $\pm$ 0.7 & 867.5 & 283.8 $\pm$ 9.8 & 12.1 $\pm$ 0.4 \\
    722.5 & 1086.0 $\pm$ 25.2 & 33.2 $\pm$ 0.8 & 872.5 & 256.5 $\pm$ 9.2 & 11.0 $\pm$ 0.4 \\
    727.5 & 1088.4 $\pm$ 25.3 & 33.6 $\pm$ 0.8 & 877.5 & 237.3 $\pm$ 9.2 & 10.3 $\pm$ 0.4 \\
    732.5 & 1158.8 $\pm$ 23.7 & 36.2 $\pm$ 0.7 & 882.5 & 229.7 $\pm$ 8.6 & 10.0 $\pm$ 0.4 \\
    737.5 & 1206.5 $\pm$ 25.1 & 38.2 $\pm$ 0.8 & 887.5 & 224.0 $\pm$ 8.1 & 9.9 $\pm$ 0.4 \\
    742.5 & 1229.9 $\pm$ 25.9 & 39.3 $\pm$ 0.8 & 892.5 & 196.1 $\pm$ 8.0 & 8.7 $\pm$ 0.4 \\
    747.5 & 1263.3 $\pm$ 27.6 & 40.9 $\pm$ 0.9 & 897.5 & 175.9 $\pm$ 7.6 & 7.9 $\pm$ 0.3 \\
    \bottomrule
	\end{tabular}
\end{table*}

 \section{Gounaris-Sakurai Fit of the Pion Form Factor}
The pion form factor $|F_\pi|^2$ is defined as
 \begin{equation}\label{eq:FF}
  |F_\pi|^2 = \frac{3s'}{\pi\alpha\beta^3_\pi(s')}\cdot\sigma^\text{dressed}(e^+e^-\to\pi^+\pi^-)  \quad , 
 \end{equation}
where $\beta_\pi = \sqrt{1-4m_\pi^2/s'}$ denotes the pion velocity. The factor $\frac{3s'}{\pi\alpha\beta^3_\pi(s')}$ from pure QED calculations is considered to be exact. Thus, the statistical error-covariance matrix of the pion form factor is constructed analogously to Eq.~\ref{eq:errprop} from the covariance matrix of the event yield, which is propagated according to Eqs.~\ref{eq:dxs} and~\ref{eq:FF} to the level of the form factor. The diagonal elements of the matrix are presented as updated statistical uncertainties of the pion form factor in Tab.~\ref{tab:xsec_FF}.
  
In the original work, a fit of the Gounaris-Sakurai parametrization~\cite{Gounaris:1968mw} to the pion form factor is used to compare the BESIII measurement to previous publications. In the fit, the statistical covariance matrix is not considered. Instead, the uncertainties before having applied the unfolding procedure are considered. These are assumed to implicitly take into account all correlations. A good fit quality is achieved, but cannot be reproduced using the originally published covariance matrix in Ref.~\cite{Colangelo:2018mtw} .
    
\begin{figure*}[htb]
  \centering
  \includegraphics[trim=0 2em 0 0, clip, width=.49\linewidth]{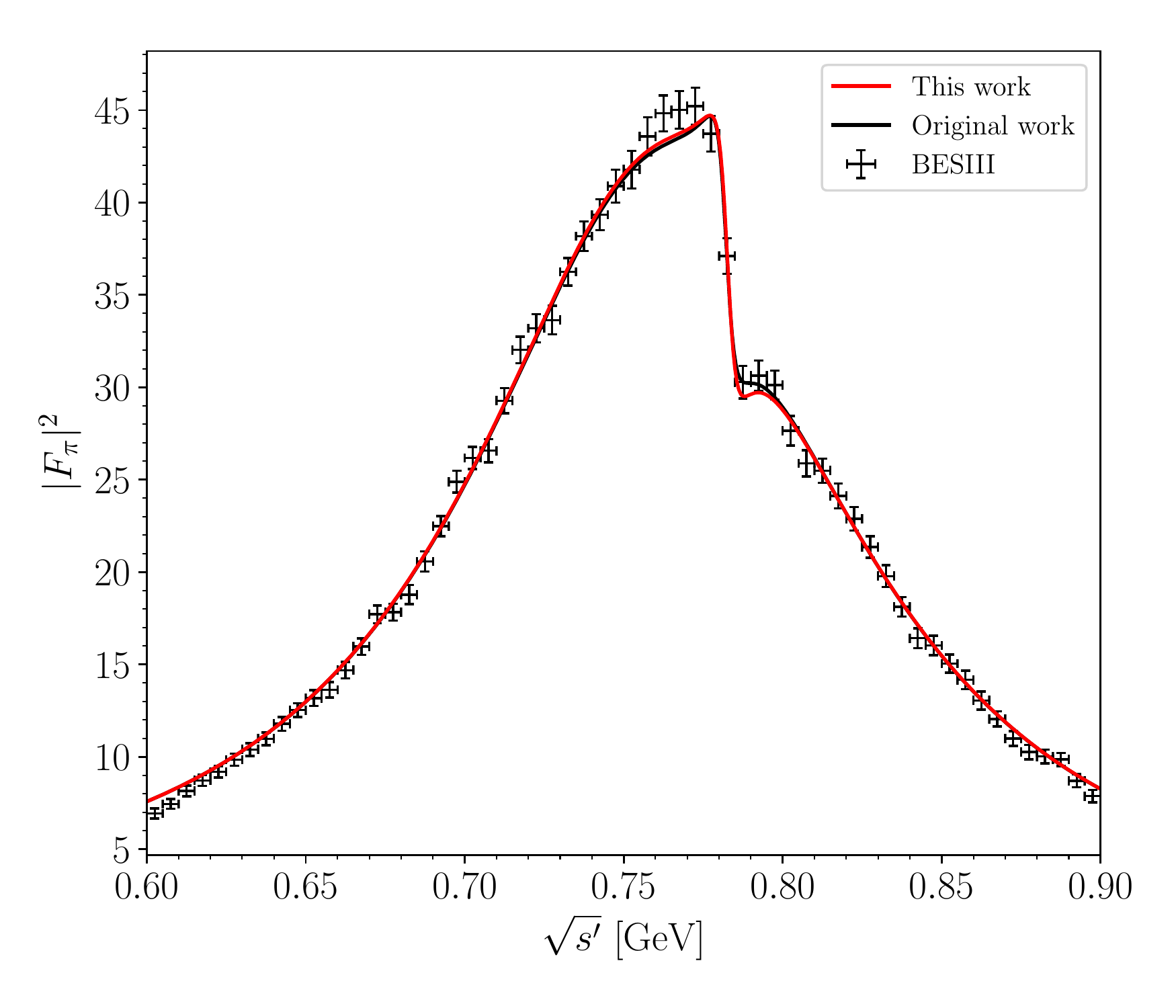}\hfill%
  \includegraphics[trim=0 2em 0 0, clip, width=.49\linewidth]{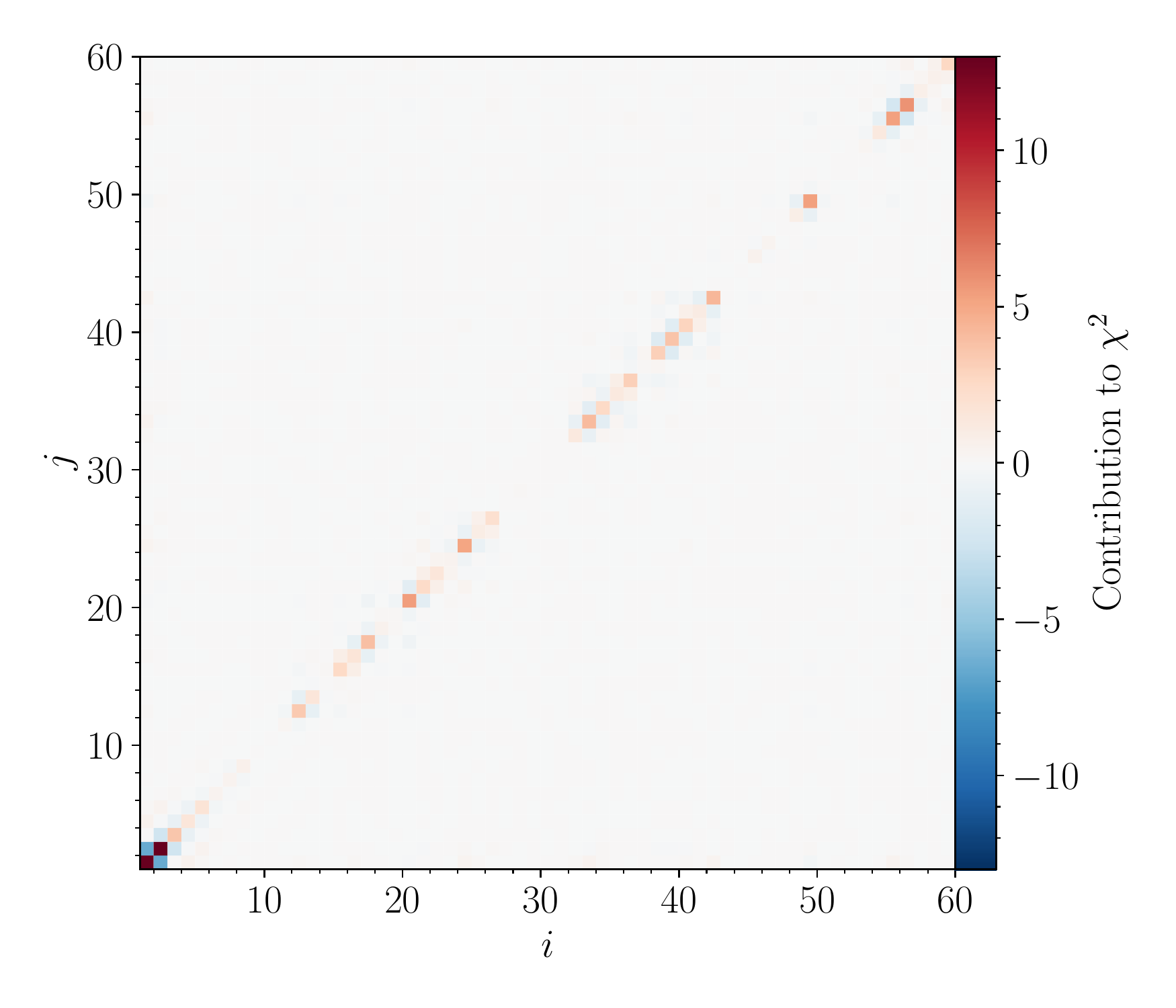}
  \includegraphics[trim=0 2em 0 0, clip, width=.75\linewidth]{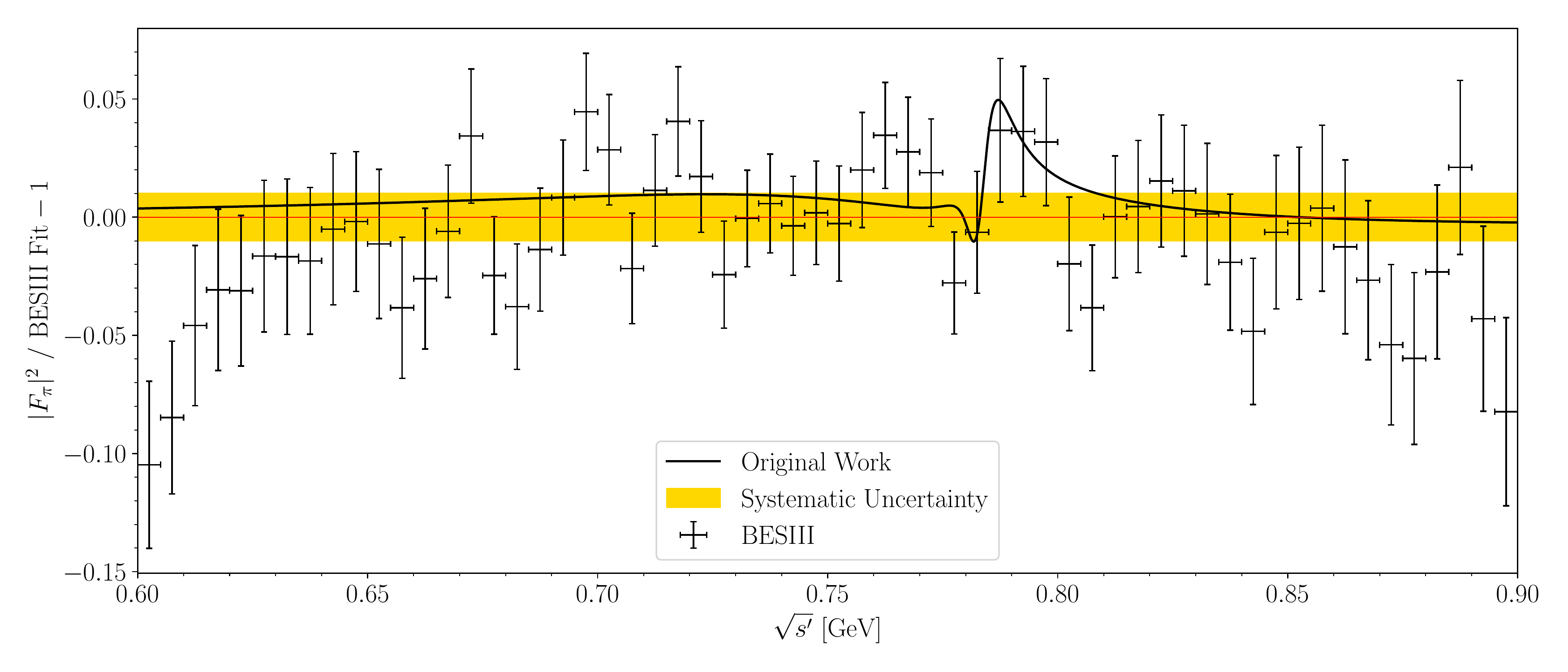}
  \caption{(Color online) {\bf Top Left:} Results of the Gounaris-Sakurai fit of the original work~\cite{Ablikim:2015orh} (black) and  this work (red); {\bf Top Right:} Contribution to the $\chi^2$ value of the individual bins of the covariance matrix; {\bf Bottom:} Deviations between the fit result of this work (red line) and the data as well as the old fit~\cite{Ablikim:2015orh} (black line).}
  \label{fig:chi2}
\end{figure*}

In this erratum, we repeat the fit of the form factor as a cross check of the newly derived covariance matrix. In order to evaluate the effects of the different treatment of the statistical errors, the width of the $\omega$ meson $\Gamma_\omega$ is fixed to the PDG value~\cite{Zyla:2020zbs}, and the masses and widths of the higher $\rho$ states $\rho(1450)$, $\rho(1700)$, and $\rho(2150)$ are fixed to the values obtained by the BaBar collaboration~\cite{Lees:2012cj}, as done in the original work. The updated fit result is illustrated with a red line in the top left panel of Fig.~\ref{fig:chi2} and compared to the original fit result. The updated fit yields a reduced $\chi^2$ value of $\chi^2 / \text{n.d.f.} = 70.70/56$. The bottom panel of Fig.~\ref{fig:chi2} illustrates the deviations of the updated fit result and the old fit result from Ref.~\cite{Ablikim:2015orh}. A clear deviation is found at the $\rho-\omega$ interference, where using the covariance matrix in the fit has worsened the agreement with data. The effect is related to fixing $\Gamma_\omega$ in the fit. However, floating the width does not only result in a better agreement between data and fit function, but also in a value of $\Gamma_\omega$ more than two standard deviations larger than the PDG value. The top right panel of Fig.~\ref{fig:chi2} shows the individual contributions of the bins of the covariance matrix to the total $\chi^2$ value. Large fluctuations, as reported by Colangelo \textit{et al.}\cite{Colangelo:2018mtw} are not observed. The largest contribution to $\chi^2$ stems from the mass region between 600 and 615\,MeV, where there is a systematic difference between the data and the Gounaris-Sakurai parametrization. The fit results are summarized in Tab.~\ref{tab:fit_results}.
\begin{table*}[htb]
  \caption{Fit results together with the statistical uncertainties from this work (BESIII), the original work (BESIII~16~\cite{Ablikim:2015orh}), the BaBar measurement~\cite{Lees:2012cj}, and the PDG values~\cite{Zyla:2020zbs}.}
  \label{tab:fit_results}
  \begin{tabular*}{\linewidth}{l@{\extracolsep{\fill}}r@{\extracolsep{\fill}}r@{\extracolsep{\fill}}r@{\extracolsep{\fill}}r}
    \toprule
    Parameter & BESIII & BESIII 16 & BaBar$\phantom{BB}$ & PDG \\
    \midrule\midrule
    $m_\rho$ [MeV]      & 776.58$\pm$0.42 & 776.0$\pm$0.4 & 775.02$\pm$0.31$\phantom 8$ & 775.26$\pm$0.25 \\
    $\Gamma_\rho$ [MeV] & 152.05$\pm$0.65 & 151.7$\pm$0.7 & 149.59$\pm$0.67$\phantom 8$ & 147.8$\pm$0.9$\phantom 0$  \\
    $m_\omega$ [MeV]    & 782.69$\pm$0.34 & 782.2$\pm$0.6 & 781.91$\pm$0.18$\phantom 8$ & 782.65$\pm$0.12 \\
    $|c_\omega|$ [$10^{-3}$] & 1.92$\pm$0.16& 1.7$\pm$0.2 & 1.644$\pm$0.061& --\\
    $\phi_\omega$ [rad]& 0.15$\pm$0.11& 0.04$\pm$0.13& -0.011$\pm$0.037& --\\	
    \midrule
    $\chi^2 / \text{n.d.f.}$ & 70.70 / 56 & 49.1 / 56 & -- & -- \\
    \bottomrule
  \end{tabular*}
\end{table*}

By comparing the resulting parameters one finds a significant improvement of the uncertainty of the $\omega$ mass. The results obtained for other parameters agree well with the original work. Thus, the systematic differences found between the BESIII result and previous measurements using the Gounaris-Sakurai fit in Ref.~\cite{Ablikim:2015orh} can be considered unchanged. The deviations between the fit results of BESIII and BaBar are on the level of 3$\sigma$ or less, which might be well covered by systematic effects that are neglected at this point. It should also be noted that the BaBar results do not consider the covariance matrix in the fit due to expected biases~\cite{Lees:2012cj}. The precise determination of resonance parameters is not the purpose of this erratum.

\section{Reevaluation of \texorpdfstring{\boldmath${a_\mu^{\pi\pi\mathrm{, LO}}}(600 - 900\,\mathrm{MeV})$}{amu(LO)(pipi)(600-900 MeV)}}
The hadronic vacuum polarization (HVP) contribution to the muon anomalous magnetic moment $a_\mu$ can be connected to the cross section $\sigma(e^+e^-\to\text{hadrons})$ using the optical theorem~\cite{Eidelman:1995ny}. The contribution of $e^+e^-\to\pi^+\pi^-$ to $a_\mu$ in the mass range of the $\rho$--$\omega$ interference is given by
 \begin{equation}
  a_\mu^{\pi\pi\mathrm{, LO}}(600 - 900\,\mathrm{MeV}) = \frac{1}{4\pi^3}\int\displaylimits_{(600\,\text{MeV})^2}^{(900\,\text{MeV})^2}\text{d}s'\, K(s')\,\sigma^\text{bare}(e^+e^-\to\pi^+\pi^-(\gamma_\text{FSR})) \quad , \label{eq:apipi}
 \end{equation}
where $K(s')$ is a kernel function.

With the systematical uncertainty remaining at 0.9\,\% \cite{Ablikim:2015orh}, the BESIII result on the hadronic vacuum polarization now reads as $a_\mu^{\pi\pi\mathrm{, LO}}(600 - 900\,\mathrm{MeV}) = (368.2 \pm 1.5_{\rm stat} \pm 3.3_{\rm syst})\times 10^{-10}$.
\begin{figure*}[hbt]
  \centering
  \includegraphics[trim=0 2em 0 1em, clip, width=.8\textwidth]{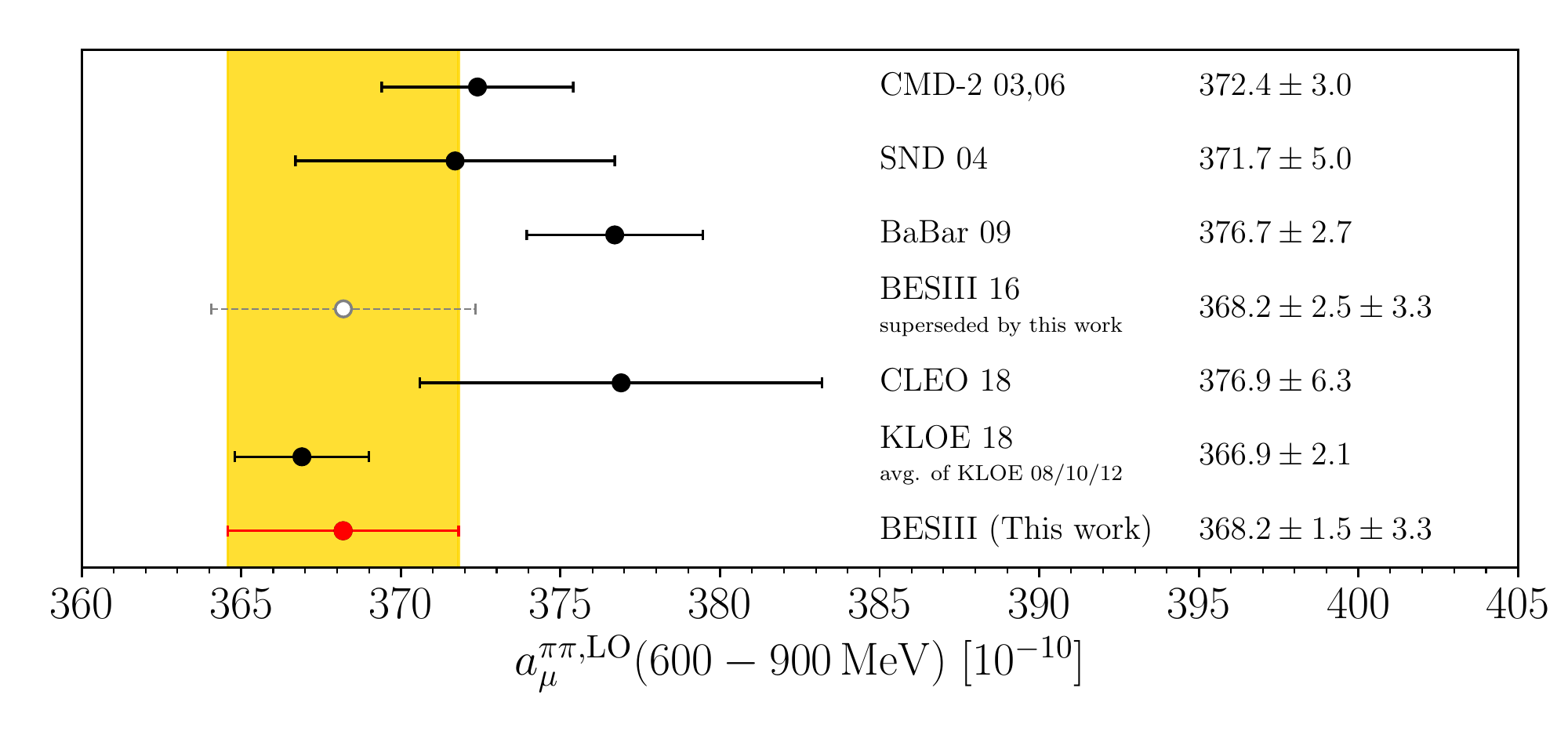}
  \caption{\label{fig:amu}(Color online) Comparison of the updated calculation of the leading-order (LO) hadronic vacuum polarization contribution to $(g-2)_\mu$ due to $\pi^+\pi^-$ in the energy range 600 - 900 MeV from BESIII and the corresponding results from CMD-2~\cite{Akhmetshin:2003zn,Akhmetshin:2006bx}, SND~\cite{Achasov:2006vp}, BaBar~\cite{Lees:2012cj}, BESIII\,16~\cite{Ablikim:2015orh},CLEO~\cite{Xiao:2017dqv}, and KLOE~\cite{Anastasi:2017eio}. The respective values are taken from the white paper of the \emph{Muon g-2 Theory Initiative}~\cite{Colangelo:2018mtw,Davier:2019can,Aoyama:2020ynm,Davier:2017zfy,Keshavarzi:2018mgv,Hoferichter:2019gzf,Keshavarzi:2019abf}. The yellow band indicates the 1$\sigma$ range of the updated BESIII result.}
\end{figure*}

Figure~\ref{fig:amu} shows the results of the calculation compared to previous measurements. The statistical uncertainty is reduced by 40\,\% compared to the original work. The result lines up well with the KLOE results, while the 1.7$\sigma$ discrepancy between the BESIII and BaBar results remains.

\section*{Acknowledgements}
The BESIII collaboration thanks the staff of BEPCII and the IHEP computing center for their strong support. This work is supported in part by National Key Basic Research Program of China under Contract No. 2015CB856700; National Natural Science Foundation of China (NSFC) under Contracts Nos. 11625523, 11635010, 11735014, 11822506, 11835012, 11935015, 11935016, 11935018, 11961141012; the Chinese Academy of Sciences (CAS) Large-Scale Scientific Facility Program; Joint Large-Scale Scientific Facility Funds of the NSFC and CAS under Contracts Nos. U1732263, U1832207; CAS Key Research Program of Frontier Sciences under Contracts Nos. QYZDJ-SSW-SLH003, QYZDJ-SSW-SLH040; 100 Talents Program of CAS; INPAC and Shanghai Key Laboratory for Particle Physics and Cosmology; ERC under Contract No. 758462; German Research Foundation DFG under Contracts Nos. 443159800, Collaborative Research Center CRC 1044, FOR 2359, FOR 2359, GRK 214; Istituto Nazionale di Fisica Nucleare, Italy; Ministry of Development of Turkey under Contract No. DPT2006K-120470; National Science and Technology fund; Olle Engkvist Foundation under Contract No. 200-0605; STFC (United Kingdom); The Knut and Alice Wallenberg Foundation (Sweden) under Contract No. 2016.0157; The Royal Society, UK under Contracts Nos. DH140054, DH160214; The Swedish Research Council; U. S. Department of Energy under Contracts Nos. DE-FG02-05ER41374, DE-SC-0012069.

\bibliography{bib}

\end{document}